\documentclass[twocolumn,prl,aps,tightenlines,floatfix]{revtex4}
\usepackage{amsfonts,amsmath,amssymb,amsthm,bm}
\usepackage{graphicx}
\usepackage[dvipsnames,usenames]{color}
\begin{document}
\bibliographystyle{apsrev}
\title{Seebeck Coefficients in Nanoscale Junctions: Effects of Electron-vibration Scattering and Local Heating.}
\author{Bailey C. Hsu $^{1}$ }
\author{Yu-Shen Liu $^{1,4}$ }
\author{Sheng Hsien Lin $^{2,3}$}
\author{Yu-Chang Chen $^1$}
\email{yuchangchen@mail.nctu.edu.tw} \affiliation{$^1$Department of
Electrophysics, National Chiao Tung University, 1001 Ta Hsueh Road,
Hsinchu 30010, Taiwan\\
$^2$Department of Applied Chemistry, National Chiao Tung University, 1001 Ta Hsueh Road,
Hsinchu 30010, Taiwan\\
$^3$ Institute of Atomic and Molecular Sciences, Academia Sinica, Taipei 106, Taiwan\\
$^4$ College of Physics and Engineering, Changshu Institute of Technology, Suzhou, 215500, China}
\begin{abstract}
We report first-principles calculations of inelastic Seebeck
coefficients in an aluminum monatomic junction. We compare the elastic and
inelastic Seebeck coefficients with and without local heating. In the
low temperature regime, the signature of normal modes in the
profiles of the inelastic Seebeck effects is salient. The inelastic Seebeck
effects are enhanced by the normal modes, and further magnified by local heating.
In the high temperature regime, the inelastic Seebeck effects
are weakly suppressed due to the quasi-ballistic transport.
\end{abstract}
\maketitle

The electron-vibration interaction plays an important role in molecular
electronics. Electrons flowing in nanojunctions are
characterized by quasi-ballistic electron transport~\cite{Rantnerreview}.
Only a small fraction of electrons experience the inelastic scattering.
Electron-vibration interactions cause discontinuities in the current-voltage
(I-V) characteristics known as the inelastic current tunneling
spectroscopy (IETS)~\cite{Kushmerick}. The IETS can provide information
on the underlying atomic structures of junctions~\cite{chenh2}. It also
gives important signals to the molecular junction characterization~\cite{Kiguchi}.

Electrons that travel with energies larger than the energy of normal
modes can excite corresponding vibrations in the nano-structure anchoring
the electrodes. This effect causes local heating in the
nano-structure~\cite{chen1,Frederiksen,Huang2}. Heating occurs when electrons
exchange energy with the excitation and relaxation of the energy levels of
the vibration of the nano-structured object that anchors
the electrodes. The heating power is typically within 15\% of the
electric power ($IV_{B}$) supplied by a battery even at ambient temperatures
because of the quasi-ballistic transport. The heat generated in the central
wire region is dissipated to the bulk electrodes via phonon-phonon interactions. The
heat generation eventually equilibrates the heat dissipation, where the
wire region reaches an effective local temperature $T_{w}$ higher than the
electrode temperature $T_{e}$. Local temperature depends on several
factors: the strength of coupling between electrons and the vibrations,
the background temperature, and the thermal current which dissipates heat.

In the last decade, remarkable progress has been achieved in measuring the
Seebeck coefficients in nanojunctions~\cite{Ruitenbeek,Majumdar1,Majumdar2,Majumdar3}.
These experiments have shed light on the design of possible energy-conversion
nano-devices, such as nanoscale refrigerators and power generators~\cite{liupowergen}.
These experiments have also inspired rapid development in the theory of
thermoelectric nanojunctions~\cite{Paulsson,Zheng,Pauly,Dubi,Ke,Finch,Bergfield,Liu2,Galperin2,Troels}.
In bulk systems, diffused electrons scattered by phonons can significantly
affect the Seebeck coefficient. However, the effects of the quasi-ballistic
electrons scattered by vibrations of the nano-structure
on the Seebeck coefficient are relatively unexplored in nanojunctions.
To the best of our knowledge only one report for inelastic Seebeck
coefficients in molecular junctions based on model calculations is
available in the literature~\cite{Galperin}. In this Letter, we
investigate inelastic Seebeck coefficients from first-principles
calculations.

Following the work of Chen, Zwolak, and Di Ventra~\cite{chen1}, the
many-body Hamiltonian of the system is $H=H_{el}+H_{vib}+H_{el-vib}$,
where $H_{el}$ is the electronic part of the Hamiltonian under adiabatic
approximations and $H_{vib}$ is the ionic part of the Hamiltonian, which can be
casted into a set of independent simple harmonic oscillators via normal
coordinates. The normal mode frequencies are $\omega_{j}$, and $H_{el-vib}$ is
a part of the Hamiltonian for electron-vibration interactions which has the form of,%
\begin{align}
H_{el-vib}  &  =\sum_{\alpha,\beta,E_{1},E_{2},j}\left(  \sum_{i,\mu}%
\sqrt{\frac{\hbar}{2M_{i}\omega_{j}}}A_{i\mu,j}J_{E_{1},E_{2}}^{i\mu
,\alpha\beta}\right) \nonumber\\
&  \cdot a_{E_{1}}^{\alpha\dag}a_{E_{2}}^{\beta}(b_{j}+b_{j}^{\dag
}),\label{elph}%
\end{align}
where $\alpha,\beta=\{L,R\}$; $M_{i}$ is the mass of the $i$-th atom;
$A_{i\mu,j}$ is a canonical transformation between normal and
Cartesian coordinates satisfying $%
{\displaystyle\sum\nolimits_{i,\mu}}
A_{i\mu,j}A_{i\mu,j^{\prime}}=\delta_{j,j^{\prime}}$; $b_{j} $ is the
annihilation operator corresponding to the $j$-th normal mode, and $a^{L(R)}$ is
the annihilation operator for electrons; the coupling
constant $J_{E_{1},E_{2}}^{i\mu,\alpha\beta}$ between electrons
and the vibration of the $i$-th atom in $\mu$ ($=x$, $y$, $z$) component
can be calculated as,%
\begin{equation}
J_{E_{1},E_{2}}^{i\mu,\alpha\beta}=\int d\mathbf{r}\int d\mathbf{K}%
_{\parallel}[\Psi_{E_{1}\mathbf{K}_{\parallel}}^{\alpha}(\mathbf{r})]^{\ast
}[\partial_{\mu}V^{ps}(\mathbf{r},\mathbf{R}_{i})\Psi_{E_{2}\mathbf{K}%
_{\parallel}}^{\beta}(\mathbf{r})],\label{couplingJ}%
\end{equation}
where $V^{ps}(\mathbf{r},\mathbf{R}_{i})$ is the pseudopotential representing
the interaction between electrons and the $i$-th ion; $\Psi_{E\mathbf{K}%
_{\parallel}}^{\alpha(=L,R)}(\mathbf{r})$ stands for the effective
single-particle wave function of the entire system corresponding to
incident electrons propagated from the left (right) electrode. These wave
functions are calculated iteratively until convergence and
self-consistency are achieved in the framework of DFT combined with the
Lippmann-Schwinger equation~\cite{Lang},%
\[
\Psi_{E\mathbf{K}_{\parallel}}^{\alpha}(\mathbf{r})=\Psi_{0,E\mathbf{K}%
_{\parallel}}^{\alpha}(\mathbf{r})+\int d\mathbf{r}_{1}\int d\mathbf{r}%
_{2}G(\mathbf{r,r}_{1})V(\mathbf{r}_{1},\mathbf{r}_{2})\Psi_{E\mathbf{K}%
_{\parallel}}^{\alpha}(\mathbf{r}),
\]
where $G$ is the Green's function of the biased bimetallic electrodes
with $V_{B}=(\mu_{R}-\mu_{L})/e$, where $\mu_{R(L)}$ is the chemical
potentials deep in the right (left) electrode, respectively; the wave function of the
bimetallic junction, $\Psi_{0,E\mathbf{K}_{\parallel}}^{\alpha}(\mathbf{r})$,
is calculated by solving a combination of the
Poisson and Schr\"{o}dinger equations until self-consistency is achieved, where
the boundary conditions are given by the electrons deep inside the biased
electrodes. The inclusion of a single molecule bridging the bimetallic electrodes
is considered as the scattering center, described by the potential $V$.
\begin{figure}
\includegraphics[width=6.50cm]{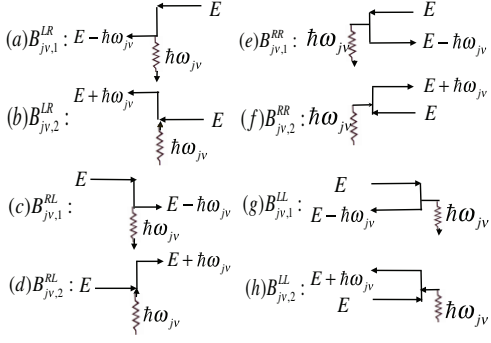}
\caption{
(color online)
Feynman diagrams of the first-order electron-vibration scattering processes
considered in this study.}
\label{Fig1}
\end{figure}

Our starting point is the inelastic current when considering electron-vibration
interactions,%
\begin{equation}
I(\mu_{L},T_{L};\mu_{R},T_{R};T_{w})=\frac{2e}{h}\int dE[(f_{E}^{R}-f_{E}%
^{L})-(\tilde{B}^{R}-\tilde{B}^{L})]\tau(E),\label{inelasticI}%
\end{equation}
where $f_{E}^{L\left(  R\right)  }=1/\{\exp[(E-\mu_{L(R)})/(k_{B}%
T_{L(R)})]+1\} $ is the Fermi-Dirac distribution function describing the
statistic of electrons deep in the left (right) electrode with
temperature $T_{L(R)}$ and chemical potential $\mu_{L(R)}$; the transmission
function $\tau(E)=\frac{\pi\hbar^{2}}{mi}\int d\mathbf{R}\int d\mathbf{K}_{||}%
(\Psi_{EK_{\parallel}}^{R\ast}\nabla\Psi_{EK_{\parallel}}^{R}-\nabla
\Psi_{EK_{\parallel}}^{R\ast}\Psi_{EK_{\parallel}}^{R})$ is calculated from
the electronic part of the wave functions $\Psi_{EK_{\parallel}}^{R}$.
The terms $\tilde{B}^{L(R)}$ represent the corrections to the elastic current
considering the eight first-order scattering processes depicted
in Fig.~\ref{Fig1}, %
\begin{equation}
\tilde{B}^{\alpha}=\sum_{j}[\langle|B_{j,k}^{\beta,\alpha}|^{2}\rangle
f_{E}^{\alpha}(1-f_{E\pm\hbar\omega_{j}}^{\beta})-\langle|B_{j,k}%
^{\alpha\alpha}|^{2}\rangle f_{E}^{\alpha}(1-f_{E\pm\hbar\omega_{j}}^{\alpha
})],\label{BLR}%
\end{equation}
where $\alpha,\beta=\{L,R\}$ and $\alpha\neq\beta$. The
parameters $B_{j,1(2)}^{RR}$ and $B_{j,1(2)}^{LR}$ denoted in Eq.~(\ref{BLR})
are,
\begin{equation}
B_{j,1(2)}^{\alpha R}=i\pi\sum_{i\mu}\sqrt{\frac{\hbar}{2\omega_{j}}}%
A_{i\mu,j}J_{E\pm\hbar\omega_{j},E}^{i\mu,\alpha R}D_{E\pm\hbar\omega_{j}%
}^{\alpha}\sqrt{\delta+\left\langle n_{j}\right\rangle },\label{BAR}%
\end{equation}
where $\alpha=\{L,R\}$; $\delta=0~(1)$ represents the process of phonon
emission (absorption); the other two parameters in Eq.~(\ref{BLR}) can be
obtained by the relations $B_{j,1(2)}^{LL}=-B_{j,1(2)}^{RR}$ and
$B_{j,1(2)}^{RL}=-B_{j,1(2)}^{LR}$; the average number of local phonons is
$\left\langle n_{j}\right\rangle =1/\{\exp[\hbar\omega_{j}/(k_{B}T_{w})]-1\}$
, where $T_{w}$ is the effective wire temperature.

The rate of energy absorbed (emitted) by the anchored nano-structures due to
incident electrons from the $\beta=\{L,R\}$ electrode and scattered to
the $\alpha=\{L,R\}$ electrode via a vibrational mode $j$ is denoted
by $W_{j}^{\alpha\beta,2(1)}$. The total thermal power generated in
the junction $P$, calculated from the Fermi golden rule, can be written
as the sum of all the vibrational modes of eight scattering processes
shown in Fig.~\ref{Fig1},%
\begin{equation}
P=\sum_{j\in vib}\sum_{\alpha=\{L,R\}}\sum_{\beta=\{L,R\}}(W_{j}^{\alpha
\beta,2}-W_{j}^{\alpha\beta,1}).\label{TotPower}%
\end{equation}
The rate of heat dissipated to electrodes via phonon-phonon interactions is
calculated using the weak link model,
\begin{equation}
J_{ph}=\frac{2\pi K^{2}}{\hbar}\int_{0}^{\infty}dEEN_{L}(E)N_{R}%
(E)[n_{L}(E)-n_{R}(E)],\label{JQ}%
\end{equation}
where $K=1.59$ eV/a$_{0}^{2}$ is the stiffness of the 4-Al atom chain
connected to the electrodes obtained from the total energy calculation~\cite{Yang};
$N_{L(R)}(E)$ is the spectral density of local phonon DOS
at the left (right) electrode surface from first-principles calculations~\cite{Chulkov}; and
$n_{L(R)}\equiv1/(e^{E/K_{B}T_{L(R)}}-1)$ is the Bose-Einstein distribution
function. The effective local temperature $T_{w}$ is obtained when heat generation in the
nano-structure and heat dissipation into the bulk electrodes reach balance.

We calculate the inelastic Seebeck coefficient based on the inelastic current
described in Eq.~(\ref{inelasticI}), which is a function of
$T_{L}$, $T_{R}$, $T_{w}$, and $V_{B}=(\mu_{R}-\mu_{L})/e$. We consider
an extra current induced by an infinitesimal temperature difference
($\Delta T$) across the junction. This current is counterbalanced by an extra
current driven by a voltage ($\Delta V$), which is induced by $\Delta T$ via
the Seebeck effect, \emph{i.e.},%
\begin{align}
I(\mu_{L},T_{L};\mu_{R},T_{R})  & =[I(\mu_{L},T_{L}-\frac{\Delta T}{2};\mu
_{R},T_{R}+\frac{\Delta T}{2})\nonumber\\
& +I(\mu_{L}-\frac{e\Delta V}{2},T_{L};\mu_{R}+\frac{e\Delta V}{2}%
,T_{R})]/2.\label{DeltaI}%
\end{align}
After expanding the above equation to the first order in $\Delta T$ and
$\Delta V$, we obtain the inelastic Seebeck coefficient (defined as
$S_{el+vib}=\Delta V/\Delta T$),%
\begin{equation}
S_{el+vib}=-\frac{1}{e}\frac{\int dE(\tilde{\frac{\partial f_{E}^{R}}{\partial
T_{R}}}+\tilde{\frac{\partial f_{E}^{L}}{\partial T_{L}}})\tau(E)}{\int
dE(\tilde{\frac{\partial f_{E}^{R}}{\partial E}}+\tilde{\frac{\partial
f_{E}^{L}}{\partial E}})\tau(E)},\label{inelasticS}%
\end{equation}
where%
\begin{equation}
\tilde{\frac{\partial f_{E}^{\alpha}}{\partial E}}=\frac{\partial
f_{E}^{\alpha}}{\partial E}-\sum_{j\in vib;k=1,2}(C_{\mu,j,k}^{R\alpha}%
+C_{\mu,j,k}^{L\alpha});\label{FRE}%
\end{equation}%
\begin{equation}
\tilde{\frac{\partial f_{E}^{\alpha}}{\partial T_{R}}}=\frac{\partial
f_{E}^{\alpha}}{\partial T_{R}}-\sum_{j\in vib;k=1,2}(C_{T,j,k}^{R\alpha
}+C_{T,j,k}^{L\alpha}),\label{FRT}%
\end{equation}
where $\alpha=\{L,R\}$ and the parameters $C_{\mu,j,1(2)}^{\alpha R}$ and
$C_{T,j,1(2)}^{\alpha R}$ are,
\begin{equation}
C_{\mu,j,1(2)}^{\alpha R}=[f_{E}^{R}\frac{\partial f_{E\pm\hbar\omega_{j\nu}%
}^{\alpha}}{\partial E}-(1-f_{E\pm\hbar\omega_{j}}^{\alpha})\frac{\partial
f_{E}^{R}}{\partial E}]\langle|B_{j,1(2)}^{RR}|^{2}\rangle;
\label{CRR1}%
\end{equation}
\begin{align}
C_{T,j,1(2)}^{\alpha R}  & =[\frac{E\pm\hbar\omega_{j}-\mu_{\alpha}}{T_{R}%
}f_{E}^{R}\frac{\partial f_{E\pm\hbar\omega_{j}}^{\alpha}}{\partial
E}\nonumber\\
& -\frac{E-\mu_{R}}{T_{R}}(1-f_{E\pm\hbar\omega_{j}}^{\alpha})\frac{\partial
f_{E}^{R}}{\partial E}]\langle|B_{j,1(2)}^{\alpha R}|^{2}\rangle,\label{CaR}%
\end{align}
where $\alpha=\{L,R\}$ and $B_{j,1(2)}^{\alpha\beta}$ are given by Eq.~(\ref{BAR}).
The other two terms in Eq.~(\ref{inelasticS}) can be calculated
with the following relations $\frac{\partial\tilde{f}_{E}^{L}}{\partial
T}=\frac{\partial\tilde{f}_{E}^{R}}{\partial T}(L\rightleftharpoons R)$ and
$\frac{\partial\tilde{f}_{E}^{L}}{\partial E}=\frac{\partial\tilde{f}_{E}^{R}%
}{\partial E}(L\rightleftharpoons R)$, where $L\rightleftharpoons R$ represents the
interchange between $R$ and $L$. We see that, in the absence of
electron-phonon scattering, Eq.~(\ref{inelasticS}) recovers the elastic Seebeck
coefficient described in Ref.~\cite{Liu}.
\begin{figure}
\includegraphics[width=7.50cm]{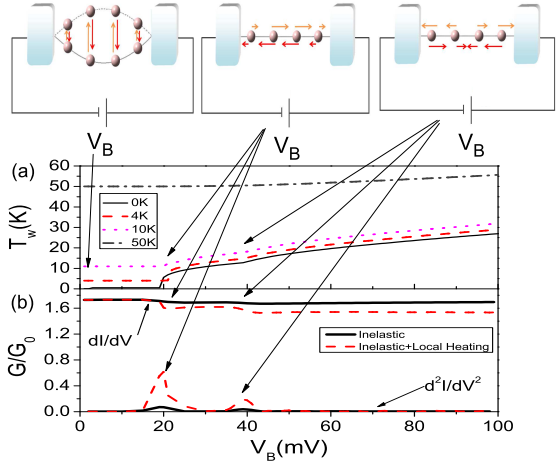}
\caption{
(color online)
(a) Local temperature $T_{w}$ as a function of $V_{B}$ for
$T_{e}=0$, $4$, $10$, $50$~K; (b) the differential conductance
and the absolute value of $dG/dV$ due the electron-vibration
interaction without [solid (black) line] and with [dash (red) line]
local heating as a function of bias for $T_{e}=12$~K. The schematic
shows the normal modes that contribute to the jumps in the
local temperature and inelastic current profiles.}
\label{Fig2}
\end{figure}

We now apply our theory to investigating the inelastic Seebeck effects
of four Al atoms bridging the bimetallic Al electrodes modeled as electron
jellium with $r_{s}\approx2$. The 4-Al junction is structurally and
electronically simple such that the first-principle calculations reported
here can be performed with a high level of accuracy. It, therefore, serves
as an ideal testbed for comparing the predictions of theory and measurements
in experiments. We compare the elastic and inelastic Seebeck coefficients
assuming that the left and right electrodes share the same
temperature $T_{e}$. In order to qualitatively show to what extent local
heating affects the inelastic Seebeck coefficient, we choose to display
inelastic Seebeck coefficients with and without local heating.
\begin{figure}
\includegraphics[width=8.50cm]{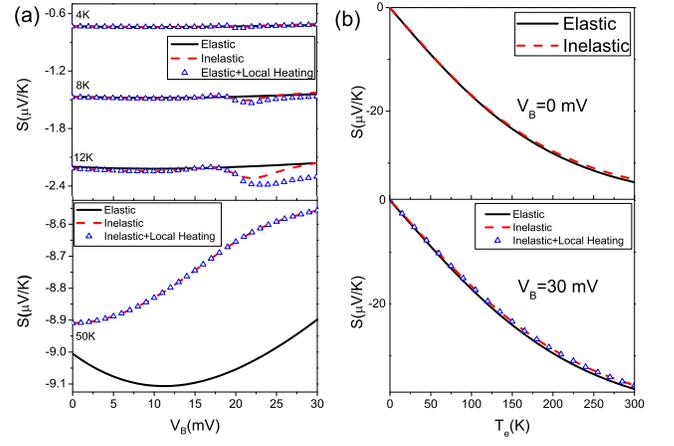}
\caption{
(color online)
Elastic Seebeck coefficient [solid (black) line], inelastic
Seebeck coefficient without local heating [dash (red) line],
and that with local heating [triangle (blue) line] (a) as a function
of bias $V_{B}$ for $T_{e}=4$, $8$, $12$ K (in upper panel) and $T_{e}=50$ K
(in lower panel); and (b) as a function of $T_{e}$ for
$V_{B}=0$ K (in upper panel) and and $V_{B}=30$ K (in lower panel.)
}
\label{Fig3}
\end{figure}

In the case of
\textquotedblleft without local heating\textquotedblright, we mean that
the heat generated in the wire region is perfectly dissipated to
electrodes such that $T_{w}=T_{e}$. When
including \textquotedblleft local heating\textquotedblright, the
effective local wire temperature $T_{w}$ is higher than the
electrode temperature $T_{e}$. Fig.~\ref{Fig2}(a) shows $T_{w}$ as a
function of the applied bias $V_{B}$ for various $T_{e}$.
We note that three jumps occur at $V_{B}=2.5$, $20$, and $40$~mV, corresponding
to the energies of the normal modes. The sharp increase in $T_{w}$
at $V_{B}=20$ mV corresponds to the first longitudinal vibrational
mode. Two degenerate transverse modes are present in the $x$- and $y$-directions
at $V_{B}=$2.5 mV, and we show the representative one in Fig.~\ref{Fig2}.
Due to the selection rule, the contributions
to local heating from modes with vibrational components perpendicular
to the direction of electron transport ($z$-direction) are unimportant.
For $T_{e}=0$, $4$, and $10$~K, $T_{w}$
displays larger jumps at $V_{B}=20$ mV, where $eV_{B}$ is
the energy of the first longitudinal vibrational mode.
For $T_{e}=50$~K, the signatures of normal modes in $T_{w}$
are wiped out by high temperatures.
The increase in local temperature is less significant at higher $T_{e}$.
This is due to increasingly efficient heat dissipation caused by the
increase of phonon population in the electrodes, as shown in Eq.~(\ref{JQ}).
Fig.~\ref{Fig2}(b) shows the inelastic profile of the conductance
($G=dI/dV$) and derivative of conductance ($d^{2}I/dV^{2}$) as a
function of bias with and without local heating. Local heating
enhances the effects of the electron-vibration interactions on the
inelastic current because of increased average number of local phonons.

Figure~\ref{Fig3}(a) shows Seebeck coefficients as a function of the applied
bias $V_{B}$ for various $T_{e}$. For each temperature, we calculate Seebeck
coefficients in three cases: elastic Seebeck coefficients $S_{0}$, inelastic
Seebeck coefficients without local heating $S_{1}$, and inelastic Seebeck coefficients
with local heating $S_{2}$.
The difference between the elastic and inelastic Seebeck effects is more salient
in the low temperature regime around $V_{B}=20$~mV [see the cases of
$4$, $8$, and $12$ K in the upper panel of Fig.~\ref{Fig3}(a)].
The profile of inelastic Seebeck coefficients vs. $V_{B}$
displays a strong signature corresponding to the longitudinal
vibrational mode at $V_{B}=20$~mV, where
the magnitude of the Seebeck coefficients are increased.
This feature is related to the suppression of the inelastic
current around $V_{B}=20$ mV [Fig.~\ref{Fig2}(b)], where the
transmission function effectively decreases. This
leads to larger magnitudes of Seebeck coefficients
because $S\propto-\tau^{\prime}(\mu)/\tau(\mu)$~\cite{Liu}.

The inclusion of local heating enhances the effect of
electron-vibration on Seebeck coefficients further. In the low temperature
regime, the upper panel of Fig.~\ref{Fig3}(a) shows that
$S_{1}$ (without local heating) significantly differs from
$S_{2}$ (with local heating). This is because of the large difference
between $T_{w}$ and $T_{e}$, as shown in Fig.~\ref{Fig2}(a).
For $V_{B}<30$~mV, $T_{w}$ and $T_{e}$ become almost identical
when the $T_{e}$ is large. Consequently, the difference between
$S_{1}$ and $S_{2}$ becomes small [see cases of $T_{e}=50$ K in the
lower panel of Fig.~\ref{Fig3}(a).]
In all cases, the transverse modes are negligible to the inelastic
Seebeck coefficients. Fig.~\ref{Fig3}(b) shows Seebeck coefficients as a
function of $T_{e}$ for $V_{B}=0$ and $30$~mV in three cases: $S_{0}$, $S_{1}$, and $S_{2}$.
In the high temperature regime ($T_{e}>50$~K), the magnitudes
of inelastic Seebeck coefficients ($S_{1}$ and $S_{2}$) are
slightly decreased compared with the elastic Seebeck coefficients ($S_{0}$)
due to small probability of electron-vibration scattering.

In summary, we investigated the elastic and inelastic Seebeck coefficients
with and without local heating in the 4-Al atomic junction using
first-principles calculations. In the low temperature regime, the signature
of normal modes in the profiles of inelastic Seebeck effects is salient.
The inelastic Seebeck effects are enhanced by electron-vibration
interactions due to the drastic suppression of the inelastic current at the
bias corresponding to the normal mode with longitudinal vibrational character.
Local heating enhances the inelastic Seebeck effects further due to
increased average number of local phonons. In the high temperature regime, the
inelastic Seebeck effects are slightly suppressed by electron-vibration
interactions due to quasi-ballistic electron transport in nanojunctions. The
signature of normal modes in inelastic Seebeck coefficients and local
temperatures is wiped out by the tail of the Fermi-Dirac distribution.

The authors thank MOE ATU, NCHC, National Center for Theoretical
Sciences(South) in Taiwan, and NSC (Taiwan) for support under Grants NSC
97-2112-M-009-011-MY3. Y. S. thank the support of NSFC under Grant No. 10947130.

\end{document}